
\input phyzzx
\baselineskip 24pt plus 1pt minus 1pt
\hfill\vbox{\hbox{UUPHY/95/16}
\hbox {November, 1995}
\hbox {hep-th/9511021}}\break
\NPrefs
\let\refmark=\NPrefmark
\def\define#1#2\par{\def#1{\Ref#1{#2}\edef#1{\noexpand\refmark{#1}}}}
\def\con#1#2\noc{\let\?=\Ref\let\<=\refmark\let\Ref=\REFS
         \let\refmark=\undefined#1\let\Ref=\REFSCON#2
         \let\Ref=\?\let\refmark=\<\refsend}


\define\EGH
T. Eguchi, P. B. Gilkey and A. J. Hanson, Phys. Rep. {\bf 66} (1980),213

\define\GIB
G. W. Gibbons and S. W. Hawking, Phys. Lett. {\bf B78} (1978), 430

\define\POPE
G. W. Gibbons and C. N. Pope, Comm. Math. Phys. {\bf 66} (1979), 267;
G. W. Gibbons and S. W. Hawking, Comm. Math. Phys. {\bf 66} (1979), 291

\define\BPST
A. A. Belavin, A. M. Polyakov, A. S. Schwarz and Y. S. Tyupkin,
Phys. Lett. {\bf B59} (1975), 85

\define\EH
T. Eguchi and A. J. Hanson, Ann. Phys. {\bf 120} (1979), 82

\define\TEH
T. Eguchi and A. J. Hanson, Phys. Lett. {\bf B74} (1978), 249

\define\HAWKING
S. W. Hawking, Phys. Lett. {\bf A60} (1977), 81

\define\GIBBP
T. Eguchi and P. G. O. Freund, Phys. Rev. Lett. {\bf 37} (1976), 1251;
G. W. Gibbons and C. N. Pope, Comm. Math. Phys. {\bf 61} (1979), 239

\define\SHAWKING
S. W. Hawking in "Euclidean quantum gravity" (World Scientific, Singapore,
1993)

\define\BUSH
T. Busher, Phys. Lett. {\bf B159} (1985), 127; Phys. Lett. {\bf B194}
(1987) 59;
A. Giveon, M. Porrati and E. Rabinovici, Phys. Rep. {\bf 244} (1994) 77;
E. Alvarez, L. Alvarez-Gaume and Y. Lozano, hep-th/9410237 and references
therein.

\define\KSD
S. Kloster, M. M. Som and A. Das, J. Math. Phys. {\bf 15} (1974), 1096;
J. D.Gegenberg and A. Das, Gen. Rel. Grav. {\bf 16} (1984) 817;
C. P. Boyer and J. D. Finley III, J. Math. Phys. {\bf 23} (1981) 1126

\define\BIANCHI
M. Bianchi, F. Fucito, G. C. Rossi and M. Martellini, Nucl. Phys.
{\bf B440} (1995), 129

\define\BAKAS
I. Bakas, Phys. Lett. {\bf B343} (1995), 103;
I. Bakas and K. Sfetsos, hep-th/9502065.

\define\PRASAD
M. K. Prasad, Phys. Lett. {\bf B83} (1979), 310

\define\HITCHIN
N. J. Hitchin, Math. Proc. Camb. Phil. Soc. {\bf 85} (1979), 465;
P. B. Kronheimer, J. Diff. Geom. {\bf 29} (1989), 665

\define\KHURI
R. R. Khuri, Nucl. Phys. {\bf B387} (1992), 315

\define\SINGH
J. Maharana and H. Singh, hep-th/9506213.

\define\KIRITSIS
E. Kiritsis, C. Kounnas and D. Luest, Int. J. Mod. Phys. {\bf A9}
(1994), 1361


\title{\bf{DUALITY AND SELF-DUAL TRIPLET SOLUTIONS IN EUCLIDEAN GRAVITY}}
\author{Swapna Mahapatra\foot{e-mail: swapna@iopb.ernet.in}}
\address{Department of Physics, Utkal University, Bhubaneswar-751004,
India.}

\abstract
Starting from the self-dual "triplet" of gravitational instanton
solutions in Euclidean gravity, we obtain the corresponding instanton
solutions in string theory by making use of the target space duality
symmetry. We show that these dual triplet solutions can be obtained
from the general dual Taub-NUT de Sitter solution through some limiting
procedure. The dual gravitational instanton solutions obtained here are
self-dual for some cases, with respect to certain isometries, but not
always.
\endpage
Gravitational instantons are the subject of much interest in
recent times. They are defined to be nonsingular, complete, positive
definite (Riemannian metric) solutions of vacuum Einstein equations or
Einstein equations with a cosmological constant term \EGH\POPE.
The existence of such solutions is important in the study of
quantum theory of gravity \SHAWKING. These are analogous to Yang-Mills
instantons \BPST, which are defined as nonsingular solutions of classical
equations in four dimensional Euclidean space. The Yang-Mills
instantons are characterized by self-dual field strengths, whereas
the gravitational instantons are normally characterized by self-dual
or anti-self-dual curvature. There are also examples of gravitational
instantons which are not self-dual, those are the Euclidean
version of Schwarzschild and Kerr metrics. The four dimensional
Riemanninan manifolds ($\cal M$, $g_{a b}$) for gravitational instantons
can be asymptotically locally Euclidean (ALE) or asymptotically locally
flat (ALF) or compact without boundary. ALF spaces are asymptotically
flat in three spatial directions and periodic in imaginary time direction.
One of the example of such spaces is the multi Taub-NUT solution of Hawking
\HAWKING. The ALE class of solutions are flat at infinity in the four
dimensional sense modulo the identification
under a discrete subgroup of $SO(4)$. The simplest nontrivial
example of ALE space is the Eguchi-Hanson solution. Multi-instanton
solutions of Gibbons and Hawking \GIB\ also fall under this class.
The complex projective space $CP^2$ is an example of compact, anti-self-
dual instanton \GIBBP, so it solves the Einstein equation with a
cosmological constant term. The other interesting example of compact
manifold is the $K3$ space, where the metric is not known explicitly. All
these solutions have locallized gravitational field, hence are not
asymptotically flat. There are two topological invariants associated
with these solutions, namely the Euler
characteristic $\chi$ and the Hirzbruch signature $\tau$, which
can be expressed as integrals of the curvature of a four dimensional
metric.

$$\eqalign{\chi &= {1\over 128\pi^2} \int R_{a b c d}\, R^{e f g h}\,
{\epsilon^{a b}}_{e f}\,{\epsilon^{c d}}_{g h}\, {\sqrt g}\, d^4 x +
surface \,terms\cr
\tau &= {1\over 96 \pi^2} \int R_{a b c d} \,{R^{ab}}_{e f} \,
\epsilon^{c d e f}\,{\sqrt g}\, d^4 x + surface\,terms\cr}\eqn\one$$

The topological invariants are also related to nuts (isolated points)
and bolts (two surfaces), which are the fixed points of the action of
one parameter isometry groups of gravitational instantons. For example,
the Euler number $\chi$ is the sum of the number of nuts, the number of
antinuts and twice the number of bolts while the signature $\tau$ is
the number of nuts minus the number of anti-nuts. ALE instantons have
been found explicitly by Gibbons and Hawking \GIB\ and
they are known implicitly through the work of Hitchin \HITCHIN, where
Penrose's twistor technique is used.

There is a fundamental "triplet" of self dual solutions in Euclidean
gravity \EH.
These are the metric of Eguchi-Hanson, self dual Euclidean Taub-NUT
metric and the Fubini-Study metric on $CP^2$.

The Eguchi-Hanson metric is given by \TEH,
$$d s^2 = {1\over {1 - {a^4\over r^4}}} d r^2 + r^2\,({\sigma_x}^2 +
{\sigma_y}^2) + r^2 (1 - {a^4\over r^4})\,{\sigma_z}^2\eqn\two$$
In terms of the Euler angles $\theta$, $\phi$ and $\psi$, the
differential one forms $\sigma_i$ are expressed as,
$$
\eqalign{\sigma_x &= {1\over 2} (\sin\psi d\theta - \sin\theta \cos\psi
d\phi),\cr
\sigma_y &= {1\over 2} (- \cos\psi d\theta - \sin\theta \sin\psi d\phi)
,\cr
\sigma_z &= {1\over 2} (d\psi + \cos\theta d\phi)\cr}\eqn\tewpr$$

So the Eguchi-Hanson metric in terms of Euler angles is given by,
$$d s^2 = {1\over {1 - {a^4\over r^4}}} d r^2 + {r^2\over 4} (1 -
{a^4\over r^4})
(d\psi + \cos\theta d\phi)^2 + {r^2\over 4} (d\theta^2 + \sin^2\theta
d\phi^2)\eqn\two$$

This metric has a single bolt, which is a removable singularity
provided $\psi$ lies in the range $0<\psi<2\pi$. The manifold
has $\chi = 2$ and signature $\tau = - 1$. The self-dual Euclidean
Taub-NUT solution of Hawking is given by,
$$
d s^2 = {1\over 4} ({r + m\over{r - m}}) d r^2 +
m^2 ({r - m\over{r + m}}) (d\psi + \cos\theta d\phi)^2 +
{1\over 4} (r^2 - m^2) (d\theta^2 + \sin^2\theta d\phi^2)\eqn\twopp$$
This metric has a single nut singularity which is again removable.
The manifold has $\chi = 1$ and signature $\tau = 0$. Both the
Eguchi-Hanson and Taub-NUT metrics are noncompact and they satisfy
Euclidean empty space Einstein equation and have self-dual
Riemann curvature, where the dual of the Riemann tensor $R_{i j k m}$
is defined as,
$$* R_{i j k m} \equiv {1\over 2}{\sqrt g}\,\epsilon_{k m r s}
{R_{i j}}^{r s}\eqn\thr$$
If the curvature tensor satisfies the condition,
$$
* R_{i j k m} = \pm R_{i j k m},\eqn\thre$$
then it is said to be self-dual or anti self-dual depending on the sign
on r.h.s.

The third member of the "triplet"
self-dual family is the Fubini-Study metric on $CP^2$. The metric is
given by \GIBBP,
$$
d s^2 = {d r^2\over{(1 + {\Lambda r^2\over 6})^2}} + {r^2\over
{4 (1 + {\Lambda r^2\over 6})^2}} (d\psi + \cos\theta d\phi)^2 +
{r^2\over {4 (1 + {\Lambda r^2\over 6})}} (d\theta^2 + \sin^2\theta d\phi^2)
\eqn\four$$
where, $\Lambda$ is the cosmological constant. So this metric satisfies
Einstein's equation with a cosmological constant term and has an
anti self-dual Weyl tensor, where the anti-self-duality condition is given
by,
$$
C_{\alpha\beta\gamma\delta} = -{1\over 2} \epsilon_{\alpha\beta\mu\nu}
{C^{\mu\nu}}_{\gamma\delta}\eqn\for$$
This manifold is compact without boundary
having $\chi = 3$ and
$\tau = 1$. The metric has a nut as well as a bolt type singularity.
These three metrics
constitute the fundamental triplet of self-dual solutions in Euclidean
gravity. All these metrics are actually derivable from a more
general three parameter
Euclidean Taub-NUT de Sitter metric through some limiting procedure \EH.

The general Taub-NUT de Sitter (TND) metric is given by,

$$d s^2 = {\rho^2 - L^2\over {4\Delta}} d\rho^2 + (\rho^2 - L^2)
(\sigma_x^2 + \sigma_y^2) + {4 L^2\Delta\over {\rho^2 - L^2}} \sigma_z^2
\eqn\five$$

where,
$$\Delta = \rho^2 - 2 M\rho + L^2 + {\Lambda\over 4} ( L^4 +
2 L^2 \rho^2 - {1\over 3} \rho^4)\eqn\fivepr$$

If we set,
$$M = L (1 + {a^4\over{8 L^4}} + {\Lambda L^2\over 3})\eqn\six$$

in the above (TND) and put $\Lambda = 0$ and then take
the limit $L \longrightarrow \infty$
with $r^2 = \rho^2 - L^2$ held fixed, then we get back our Eguchi-Hanson
(E-H) metric. By putting $\Lambda = 0$ and $M = L$ in TND, we get the
self-dual Taub-NUT (TN) metric.
$CP^2$ is also derivable by setting \GIBBP,

$$M = L (1 + {1\over 3} \Lambda L^2)\eqn\si$$
This ensures that the metric has a right (or left) flat Weyl tensor.
One then takes the limit
$L \rightarrow \infty$ and introduces a new radial coordinate $r$,
defined by,
$$\rho^2 - L^2 = {r^2\over{(1 + {\Lambda\over 6}r^2)}}\eqn\sev$$
It is possible to extend the TND space having four bolt type singularities
to a complete, nonsingular, Riemanninan space with one nut and one bolt,
only by taking the singular limit as $L\longrightarrow \infty$.

These triplet of self-dual solutions can be regarded as special cases
of string analogue of gravitational instanton backgrounds with dilaton
$\Phi = 0$ and anti-symmetric tensor field $B_{\mu\nu} = 0$. New solutions
in string theory can be found by performing a $T$-duality transformation
on the pure gravitational instanton solutions \BUSH.

The original and the dual backgrounds are related through the following
expression,
$$
\eqalign{\tilde G_{\tau\tau}&={1\over{G_{\tau\tau}}},\cr
\tilde G_{\tau i}&={B_{\tau i}\over{G_{\tau\tau}}},\cr
\tilde G_{i j}&=G_{i j} - {G_{\tau i } G_{\tau j} - B_{\tau i} B_{\tau j}
\over{G_{\tau\tau}}}\cr}
\qquad\eqalign{\tilde B_{\tau i}&={G_{\tau i}\over{G_{\tau\tau}}},\cr
\tilde B_{i j}&=B_{i j} - {{G_{\tau i} B_{\tau j} - G_{\tau j} B_{\tau i}}
\over{G_{\tau\tau}}}.\cr}\eqn\sevn$$
$$\tilde\Phi =\Phi - {1\over 2}\log\,G_{\tau\tau}\eqn\seven$$

We can see from expression \five\ that the components of the
general Taub-NUT de Sitter metric
are independent of $\psi$ or $\phi$ coordinates. We now write down the
corresponding new solutions in string theory by using the isometry of the
original Taub-NUT solution as well as the triplet solutions.
We explicitly check that the T-dual of the
fundamental self-dual triplet solutions are again obtained from the dual
Taub-NUT de Sitter solution through a similar singular limiting procedure.
We then discuss
about the self-duality of the new solutions.

Using the isometry in the $\psi$ direction, the dual of Taub-NUT de Sitter
solution is obtained to be,

$$\eqalign{d{\tilde s_{TND}}^2&= {\rho^2 - L^2\over{4\Delta}} d\rho^2 +
{{\rho^2 - L^2}\over{L^2\Delta}} d\psi^2 + {1\over 4} (\rho^2 - L^2)
(d\theta^2 + \sin^2\theta d\phi^2);\cr
\tilde B_{\psi\phi}&= \cos\theta;\cr
\tilde{\Phi}&=-{1\over 2} \log({\L^2\Delta\over{\rho^2 - L^2}})\cr}\eqn\eit$$

Next, we shall write down the $T$-dual of Eguchi-Hanson, self-dual Taub-NUT
and Fubini-Study metric on $CP^2$ using the isometry in $\psi$ direction
and check that they too can be
obtained from the dual TND solution through the limiting procedure as
in the pure gravity case.

T-dual of the self-dual Taub-NUT solution is given by,
$$\eqalign{d{\tilde s_{TN}}^2&={1\over 4}{{\rho + m}\over{\rho -m}}
d\rho^2 +
{{\rho + m}\over{m^2(\rho - m)}} d\psi^2 + {1\over 4} (\rho^2 - m^2)
(d\theta^2 + \sin^2\theta d\phi^2);\cr
\tilde B_{\psi\phi}&=\cos\theta;\cr
\tilde\Phi&=-{1\over 2} \log[{m^2(\rho - m)\over{\rho + m}}].\cr}\eqn\nine$$

$T$-dual of Eguchi-Hanson is given by,
$$
\eqalign{d{\tilde s_{E-H}}^2&={1\over{1 -{a^4\over r^4}}} d r^2 + {4\over{r^2(
1 - {a^4\over r^4})}} d\psi^2 + {r^2\over 4} (d\theta^2 +
\sin^2\theta d\phi^2);\cr
\tilde B_{\psi\phi}&=\cos\theta;\cr
\tilde\Phi&=-{1\over 2} \log[{r^2\over 4}(1 - {a^4\over r^4})]\cr}\eqn\ten$$

Finally, the $T$-dual of $CP^2$ is given by,
$$\eqalign{d{\tilde s_{CP^2}}^2 &={d r^2\over{(1 + {\Lambda r^2\over 6})^2}} +
{{4 (1 + {\Lambda r^2\over 6})^2}\over r^2} d\psi^2 + {r^2\over
{4(1 + {\Lambda r^2\over 6})}} (d\theta^2 + \sin^2\theta d\phi^2);\cr
\tilde B_{\psi\phi} &=\cos\theta;\cr
\tilde\Phi &=-{1\over 2}\log [{r^2\over{4(1 + {\Lambda r^2\over 6})^2}}].\cr}
\eqn\eleven$$

These are the new gravitational instanton solutions in string theory
and all of them are diagonal.
These solutions satisfy the string back
ground equations of motion derived from the four dimensional low energy
effective action. Duality transformation on the $CP^2$ solution
with constant dilaton and nonzero gauge field has been discussed in ref.
\SINGH. Now the dual E-H solution
is again obtained from the dual TND solution by setting,

$$\Lambda = 0; \qquad M = L (1 + {a^4\over{8 L^4}} + {\Lambda L^2\over 3})
\eqn\ele$$
and taking the singular limit $L \longrightarrow\infty$ with $r^2 = \rho^2
- L^2$ held fixed. Dual of self-dual Taub-NUT solution is obtained by
setting $\Lambda = 0$ and $M = L$. Finally, $T$-dual of $CP^2$ is obtained
from the dual TND solution by setting
$M = L (1 + {\Lambda L^2\over 3})$ and taking the singular limit
$L\longrightarrow\infty$ with $\rho^2 - L^2 = {r^2\over{1 +
{\Lambda r^2\over 6}}}$
held fixed. The component of the anisymmetric tensor field is same in all
the three solutions as well as in the T-dual TND solution.
Dilaton $\Phi$ is also derived from the dual TND solution
through the singular limit procedure. For example,
to obtain the dilaton field $\Phi$ in the dual $CP^2$ theory from the
dual TND solution, we again have to take $L\rightarrow\infty$
limit with $\rho^2 - L^2 = {r^2\over{(1 + {\Lambda r^2\over 6})}}$
held fixed.

A simple calculation shows that for $CP^2$, as $L\longrightarrow\infty$
and with the new radial coordinate $r$,
$$L^2\Delta \longrightarrow {r^4\over{4 (1 + {\Lambda r^2\over 6})^3}}
\eqn\twelv$$
This gives,
$$\eqalign{\tilde\Phi_{TND}&= -{1\over 2}\log({L^2\Delta\over
{\rho^2 - L^2}})\cr
&\buildrel L\longrightarrow\infty \over \longrightarrow  -{1\over 2}
\log[{r^2\over{4 (1 + {\Lambda r^2\over 6})^2}}]\cr
&=\tilde\Phi_{CP^2}\cr}\eqn\thirt$$

Similarly, for Eguchi-Hanson we can express $\Delta$ as an expansion
in ${1\over L^2}$,
$$\Delta = {r^4\over{4 L^2}} (1 - {a^4\over r^4}) + o({1\over{L^4}})+
o({1\over{L^6}}) + \ldots$$
Therefore,
$$
\eqalign{\tilde\Phi_{TND}&\buildrel L\longrightarrow\infty \over
\longrightarrow -{1\over 2} \log({r^2\over 4}
(1 - {a^4\over r^4}))\cr
&=\tilde\Phi_{E-H}\cr}\eqn\fort$$

Unlike the triplet solution in Euclidean gravity, not all these dual
triplet solutions are self-dual, which can be verified from the
self-duality condition. The thing to note here is that, we
have written down all these new dual solutions in string frame.
The corresponding solutions in Einstein frame can be obtained by a
conformal transformation involving the dilaton field. The metrics in two
different frames are related in the following fashion:
$$G_{\mu\nu}^E = e^{- 2\Phi}\,G_{\mu\nu}^{\sigma}\eqn\fortone$$

The dual E-H metric in Einstein frame is given by,
$$
ds^2 = d\psi^2 + {r^2\over 4} dr^2 + {1\over 16} r^4 (1 -{a^4\over r^4})
(d\theta^2 + \sin^2\theta d\phi^2)\eqn\forttwo$$

But we find that this metic is not self-dual as the Riemann tensor
does not satisfy the self-duality condition given previously. On the
otherhand, if we
use the isometry in the $\phi$ direction, then the dual Eguchi-Hanson
solution is given by,

$$
\eqalign{d{\tilde s_{E-H}}^2 &= {4\over{r^2(1 - {a^4\over r^4} \cos^2\theta)}}
d\phi^2 + {dr^2\over{1 -
{a^4\over r^4}}} + {r^2\over 4} d\theta^2 + {{r^2 \sin^2\theta
(1 - {a^4\over r^4})}
\over{4 (1 -{a^4\over r^4} \cos^2\theta)}} d\psi^2 \cr
{\tilde B}_{\phi\psi} &= {{(1 - {a^4\over r^4})}\cos\theta\over{(1 -
{a^4\over r^4} \cos^2\theta)}};\cr
\tilde\Phi &= - {1\over 2} \log [{r^2\over 4} (1 - {a^4\over r^4}
\cos^2\theta)]\cr}\eqn\fortyfor$$

This dual metric is again obtained as a limit
of the dual TND solution (where the duality
transformation is performed w.r.t. the $\phi$ isometry) and the resulting
expression is given by,

$$
\eqalign{d{\tilde s_{TND}}^2 &= {{\rho^2 - L^2}\over {4\Delta}} d\rho^2 +
{{4 (\rho^2 - L^2)}\over{(\rho^2 - L^2)^2 \sin^2\theta + 4 L^2\Delta\cos^2
\theta}} d\phi^2 + {1\over 4} (\rho^2 - L^2) d\theta^2 +\cr
&{{L^2\Delta (\rho^2 - L^2)\sin^2\theta}\over{(\rho^2 - L^2)^2 \sin^2\theta +
4 L^2\Delta\cos^2\theta}} d\psi^2;\cr
\tilde B_{\phi\psi} &= {{4 L^2\Delta\cos\theta}\over{(\rho^2 - L^2)^2
\sin^2\theta + 4 L^2\Delta\cos^2\theta}};\cr
\tilde\Phi &= - {1\over 2}\log[{{(\rho^2 - L^2)^2 \sin^2\theta +
4 L^2\Delta\cos^2\theta}\over{4 (\rho^2 - L^2)}}].\cr}\eqn\fortyfiv$$

We check that the antisymmetric tensor field
and dilaton are also obtained from the above metric in the singular
We then transform the
dual E-H metric to Einstein frame, which is given by,

$$
d{\tilde s_{E-H}}^2 = d\phi^2 + {{r^2 (1 - {a^4\over r^4} \cos^2\theta)}\over
{4 (1 - {a^4\over r^4})}} dr^2 + {r^4\over 16} (1 - {a^4\over r^4}
\cos^2\theta)
d\theta^2 + {1\over 16} r^4 \sin^2\theta (1 -{a^4\over r^4}) d\psi^2
\eqn\fortsix$$

This metric is Ricci flat and the self-duality condition is
satisfied, moreover the killing vector field is self-dual. On the otherhand,
the self-duality condition is violated if we use the $\psi$ isometry
to obtain the dual E-H solution. This is also true because the original
E-H metric belongs to the KSD subclass \KSD, where the metric is determined
completely in terms of a single scalar field which satisfies the three
dimensional Euclidean Laplace equation. This happens only w.r.t the
killing vecor ${\partial\over{\partial\phi}}$, not w.r.t.
${\partial\over{\partial\psi}}$.
The story for the dual Taub-NUT solution is opposite. Here the
solution obtained by using the $\psi$ isometry, turns out to be self-dual.
Again this is valid when we transform the metric to Einstein frame, as in
the Eguchi-Hanson case. On the otherhand, the solution obtained by using
the isometry in $\phi$ direction is given by (in Einstein frame),

$$
\eqalign{d\tilde s^2 &= d\phi^2 + {1\over 16} [(r + m)^2 \sin^2\theta + 4 m^2
\cos^2\theta] dr^2 +
{1\over 16} [(r^2 -m^2)^2\sin^2\theta +\cr
&4 m^2 (r - m)^2\cos^2\theta] d\theta^2 +
{1\over 4} m^2 (r - m)^2\sin^2\theta d\psi^2\cr}\eqn\fortysevn$$
and this solution is not self-dual.

The dual $CP^2$ solution w.r.t. $\phi$ isometry is given by,
$$\eqalign{d{\tilde s_{CP^2}}^2 &= {{4(1 + {\Lambda\over 6}r^2)}\over
{r^2 (1 + {\Lambda\over 6} r^2 \sin^2\theta)}} d\phi^2 + {1\over{(1 +
{\Lambda\over 6}r^2)^2}} dr^2 + {r^2\over{4 (1 + {\Lambda\over 6} r^2}}
d\theta^2 +\cr
&{r^2\sin^2\theta\over{4 (1 + {\Lambda\over 6} r^2) ( 1 +
{\Lambda\over 6} r^2\sin^2\theta)}} d\psi^2 ;\cr
\tilde B_{\psi\phi} &= {\cos\theta\over{(1 + {\Lambda\over 6} r^2
\sin^2\theta)}} ;\cr
\tilde\Phi &= -{1\over 2}\log[{{r^2(1 + {\Lambda\over 6}r^2
\sin^2\theta)}\over{4 {(1 + {\Lambda\over 6}r^2)}^2}}]\cr}\eqn\forteight$$
This solution as well as the one obtained by using the $\phi$ isometry
are not anti-self-dual.

The originl self-dual solutions (E-H and Taub-NUT) we considered here
are actually the Gibbons-Hawking multi-center metric, given by the
expression,
$$
ds^2 = V^{-1} ({\bf x}) (d\tau + \omega\cdot d{\bf x}^2)^2 +
V({\bf x}) d{\bf x} \cdot d{\bf x}\eqn\fortyeit$$
where, $d{\bf x}\cdot d{\bf x}$ denotes the three dimensional Euclidean
metric and $V$ and ${\bf\omega}$ are related by,
$$\nabla V = \nabla \times {\bf\omega}\eqn\fortynin$$
This condition ensures the curvature is self-dual. Actually this
implies that $V$ satisfies the 3 dimensional Laplace equation and its
solution determines the metric completely. The most general form of the
solution is given by,
$$
V = \epsilon + \sum_{i =1}^n {m_i\over\mid x - {\bf x}_i\mid}
\eqn\fifty$$
where, $\epsilon$ and $m_i$ are arbitrary paramers. We take all
the $m_i$s to be identical, so that the singularity of the four
dimensional manifold is removable. For $\epsilon = 1$, one obtains
the multi-Taub-NUT metrics, where the self-dual Taub-NUT solution
we have considered here, corresponds to the $n =1$ case. The
choice $\epsilon = 0$ corresponds to the multi-centre Gibbons-Hawking
metric, where
$n = 1$ corresponds to flat space and $n = 2$ corresponds to the
2-center Gibbons-Hawking metric, which is equivalent to the Eguchi-Hanson
metric through a coordinate transformation
given by Prasad \PRASAD. The dual of the multi-center metric is
actually a
conformally flat metric given by (where the killing vector is ${\partial
\over{\partial\tau}}$),
$$
\eqalign{d\tilde s^2 &= V({\bf x}) (d\tau^2 + d X^2 + d Y^2 + d Z^2)\cr
\tilde B_{\tau i} &= \omega_i\cr
\tilde\Phi &= {1\over 2}\log V\cr}
\eqn\fifone$$
The conformal factor is $e^{2\Phi}$ and hence the corresponding
Einstein metric is flat. This is what we observed for the Eguchi-Hanson
case as well as in the Taub-NUT case (in Eguchi-Hanson case, the
coordinate transformation interchanges the role of $\psi$ and
$\phi$ and isometry in $\tau$ direction is same as isometry
in $\phi$ direction). Also it has been noticed before \BIANCHI\
that the dual of multi-Taub-NUT solution is same as the multi-monopole
solutions obtained by Khuri \KHURI. So the multi-monopole solutions
can also be derived from the dual TND solution in a particular
singular limit. $CP^2$ metric does
not belong to the multi-centre Gibbons-Hawking type of ansatz as
the manifold is compact, hence the metric is not obtained from the
solution of the three dimensional Laplace equation.
The isometry here is a combination of the socalled "translational"
and "rotational" killing symmetry. It has been shown in ref.\BAKAS
that the ALE instantons and the multi Taub-NUT instantons are related
through a combination of $T-S-T$ duality transformation (more precisely
through Ehlers transformation) and the
corresponding solutions are self-dual w.r.t. the translational isometry,
but not w.r.t. the rotational isometry. A class of axionic instanton
solutions and their supersymmetric extensions have been discussed
in ref.\KIRITSIS where under $T$-duality, certain
hyper-Kahler metrics
which are solutions of the Laplace equation are mapped to
quasi-Kahler backgrounds satisfying the continual toda equations.

In this paper, we have investigated the self-dual "triplet" solutions in
pure gravity. Using the target space duality symmetry in string theory,
we have obtained new gravitational instanton solutions which are
the dual Eguchi-Hanson, self-dual Taub-NUT and $CP^2$ solutions
and they are consistent backgrounds for string propagation. We show
that these dual triplet solutions can be obtained from the dual Taub-NUT
de Sitter solution through a limiting procedure analogous to the pure
gravity case. We also observe that the self-duality condition for the dual
solutions depends on the particular isometry of the original metric.
Out of the three dual triplet solutions,
E-H and Taub-NUT solutions are found to be self-dual in Einstein frame,
whereas the dual $CP^2$ solution we have obtained
here is not self-dual w.r.t. any of the isometries of the original metric.
One also obtains the dual Schwarzschild
de Sitter solution as a limit of the dual TND, where one gets the
standard metric on $S^2\times S^2$, which is compact, but does not
satisfy the half-flat condition.

{\bf ACKNOWLEDGEMENTS:}
I would like to thank Mehta Research Institute for its warm hospitality,
where part of this work was done. I thank Ashoke Sen for some comments.
A pool grant (no. 13-6764-A/94) from C. S. I. R. is acknowledged.
\refout
\end